\documentclass[conference]{IEEEtran}
\IEEEoverridecommandlockouts
\usepackage{cite}
\usepackage{amsmath,amssymb,amsfonts}
\usepackage{algorithmic}
\usepackage{graphicx}
\usepackage{textcomp}
\usepackage[table,xcdraw]{xcolor}

\usepackage{times}  
\usepackage[hyphens]{url}  
\def\BibTeX{{\rm B\kern-.05em{\sc i\kern-.025em b}\kern-.08em
    T\kern-.1667em\lower.7ex\hbox{E}\kern-.125emX}}
\begin{document}

\title{Heart Rate Estimation from Face Videos for Student Assessment: Experiments on edBB
}

\author{Javier Hernandez-Ortega, Roberto Daza, Aythami Morales, Julian Fierrez, Ruben Tolosana\\
School of Engineering, Universidad Autonoma de Madrid, Spain\\ 
{\small \texttt{\{javier.hernandezo, roberto.daza, aythami.morales, julian.fierrez, ruben.tolosana\}@uam.es}}\\
}

\maketitle

\begin{abstract}
In this study we estimate the heart rate from face videos for student assessment. This information could be very valuable to track their status along time and also to estimate other data such as their attention level or the presence of stress that may be caused by cheating attempts. The recent edBBplat, a platform for student behavior modelling in remote education, is considered in this study\footnote{https://github.com/BiDAlab/edBB}. This platform permits to capture several signals from a set of sensors that capture biometric and behavioral data: RGB and near infrared cameras, microphone, EEG band, mouse, smartwatch, and keyboard, among others. In the experimental framework of this study, we focus on the RGB and near-infrared video sequences for performing heart rate estimation applying remote photoplethysmography techniques. The experiments include behavioral and physiological data from 25 different students completing a collection of tasks related to e-learning. Our proposed face heart rate estimation approach is compared with the heart rate provided by the smartwatch, achieving very promising results for its future deployment in e-learning applications.

\end{abstract}

\begin{IEEEkeywords}
Remote Plethysmography, Heart Rate, Student Monitoring, MOOC, Behavioral Biometrics, edBB
\end{IEEEkeywords}

\section{Introduction}
\label{sec:intro}

Nowadays e-learning is experiencing a period of high growth thanks to the flexibility it provides to students who do not have the possibility to access to traditional education, like users with an employ, geographical limitations, or any other special conditions. Trying to reach that increasing market of potential students, most of higher education institutions like Stanford, Harvard, Oxford, and the MIT have started to offer new options of virtual education \cite{bowers2015students}. Moreover, episodes such as the COVID-19 outbreak in 2020 and the social distancing imposed, have demonstrated the necessity to develop new technologies to improve e-learning platforms. 

Even though e-learning presents many advantages, it also has some drawbacks, being one of the more relevant the difficulty to demonstrate if an online evaluation is really being carried out by a specific student. Without this verification step, it is hard to know if a student has acquired the knowledge associated to a certain course, or if he is incurring in some type of fraud/cheating on the evaluation, e.g. asking another person to complete his/her exam.

Biometric technologies seem to be a perfect choice to enhance virtual education environments. These technologies allow to identify a person by their physiological and behavioral characteristics, rather than traditional methods such as a password or an ID card that could be lost, forgotten, or used by another person to perform student impersonation \cite{hadid15SPMspoofing}.

The interaction between the students and the computer or the device in which they are accessing to the educational contents can be used to acquire other information about their state, e.g. their heart rate, their level of attention, and how much stressed they are \cite{allen2007photoplethysmography}. These type of factors, i.e. stress, emotional state, motivation, focus, and attention, can affect the effectiveness of the learning process \cite{pekrun2002,tresize2017}. A student who is affected by any external agent or emotion will not take as much benefit of the lessons as another that is totally focused. Traditional education theory has been centered in how to explain the contents to the students in the best way possible, but usually without considering these context and human factors. For online education, these elements are specially crucial.

The main contributions of this study are: 

\begin{itemize}
    \item A brief survey of state-of-the-art biometric and behavioral technologies based on Human-Computer Interaction (HCI) with potential application to student monitoring.
    
    
        \item The acquisition of a dataset consisting of biometrics and behavioral data using the student monitoring platform for e-learning edBBplat \cite{edBB2020_AI4EDU}. This database (edBBdb) is publicly available for research purposes (see footnote on this page).
    
    \item An experimental evaluation of heart rate estimation in the edBB framework, and the development of a baseline algorithm for heart rate estimation based on remote photoplethysmography.
    
    \item Application of the developed baseline algorithm to two different scenarios in a simulated e-learning environment: one of them consists in estimating the mean heart rate of the students over a whole session and the other consists in making a continuous heart rate estimation during a session (useful for detecting heart rate alterations).
    
\end{itemize} 

The rest of this paper is organized as follows. Section \ref{behavioral} introduces behavioral biometrics and their application to e-learning scenarios. Section \ref{system} provides details about the structure of edBBplat. Section \ref{challenges} explains the different challenges related to student monitoring proposed in the edBB framework, being one of them heart rate estimation. Section \ref{estimation} shows the experimental protocol and the results achieved for the heart rate estimation sub-challenges. Finally, conclusions are drawn in Section \ref{conclusion}.


\pagebreak

\section{Behavioral Biometrics for \\Student Monitoring based on HCI}
\label{behavioral}

Historically, the first approaches for monitoring student evaluations in remote learning have consisted in installing a special software in the student's computer. This software is intended to be connected to an institutional server in which a Learning Management System (LMS) controls that users do not perform any forbidden action during their evaluations, i.e. executing certain applications such as the web browser, making screenshots, running certain commands, etc.

The usage of online supervisors, i.e. people that manually supervise each session by webcam, allows to monitor students in real time in a similar way as in a classroom.  However, this method is not scalable to a large number of students. 


The possibilities of biometric-based technologies for monitoring online evaluations have been recently showed in real world applications like the Coursera e-learning platform. In this case, the programmers used keystroke methods \cite{morales2016keystroke,REf_TypeNet} for verifying the identity of the students enrolled in a course. 


Behavioral biometrics refers to those biometric traits that describe the way that users perform different actions \cite{jain201650}. Behavioral biometrics traits can be extracted from Human-Computer Interaction, in which a person interacts with some devices, such as computers and smartphones, in a manner that can be highly different among them \cite{salah2013understanding,shrobe2018behavioral,REf_HCI}. 

A machine learning algorithm can learn patterns from HCI data. These patterns will be affected by several factors like the acquisition sensors, the tasks that are being captured, or the human condition and behavior. Modelling these data (that usually comes from heterogeneous sources) is useful for a multitude of applications such as elearning, security, entertainment, and health.


\begin{table}[t]
\begin{center}
\caption{\textbf{Sensors and data} captured in \textbf{edBBplat} \cite{edBB2020_AI4EDU}}.
\label{tab:sensors}
\resizebox{\columnwidth}{!}{

\begin{tabular}{|c|c|c|}
\cline{1-3}
\cellcolor[HTML]{C0C0C0}{\color[HTML]{000000} \textbf{Information Type}} & \cellcolor[HTML]{C0C0C0}{\color[HTML]{000000} \textbf{Sensors}}                                                         & \cellcolor[HTML]{C0C0C0}{\color[HTML]{000000} \textbf{Sampling Rate}} \\ 

\cline{1-3}
\textbf{Video} & \begin{tabular}[c]{@{}c@{}} 4 RGB cameras\\ 2 Infrared cameras\\ 1 Depth camera\end{tabular} & 20 Hz - 30 Hz \\ 

\cline{1-3}
\textbf{\begin{tabular}[c]{@{}c@{}}Desktop Video\end{tabular}}  & Screen  & \begin{tabular}[c]{@{}c@{}}1 Hz\end{tabular}  \\ 

\cline{1-3}
\textbf{Audio} & Microphone  & \begin{tabular}[c]{@{}c@{}}8000 Hz\end{tabular}   \\ 

\cline{1-3}
\textbf{Keystroke}    & Keyboard  & 12 Hz  \\ 

\hline
\textbf{\begin{tabular}[c]{@{}c@{}}Mouse Dynamics\end{tabular}} & Mouse  & 895 Hz  \\ 

\hline
\textbf{EEG} & Band  & \begin{tabular}[c]{@{}c@{}}1 Hz\end{tabular} \\ 

\hline
\textbf{\begin{tabular}[c]{@{}c@{}}Pulse and Inertial\end{tabular}} & \begin{tabular}[c]{@{}c@{}}SmartWatch: \\PPG, Gyroscope\\ Accelerometer, Magnetometer\end{tabular} & \begin{tabular}[c]{@{}c@{}} 1 - 200 Hz\end{tabular}  \\ 

\hline
\textbf{Context Data}   & \begin{tabular}[c]{@{}c@{}}Student, Computer, Server\end{tabular}  & NA \\ 

\hline
\end{tabular}
}
\end{center}
\end{table}

Behavioral biometrics is composed by different traits like touchpad interaction~\cite{2018_TIFS_Swipe_Fierrez}, keystroking~\cite{REf_TypeNet}, mouse dynamics \cite{chen2001can,REF_Mouse}, handwriting patterns \cite{2017_PLOSONE_eBioSign_Tolosana}, and stylometry. Relevant works in this field of research demonstrate that the information coming from HCI can be used not only for user authentication, but also for characterizing other human features like~\cite{REf_HCI}: neuromotor and cognitive abilities~\cite{2018_IETB_DetectChildTouch_Acien}, physiological signals such as human pulse~\cite{REF_HRcomp}, and human behaviors/routines. 


\section{edBBplat: A Platform of Biometrics and Behavior for Remote Education}
\label{system}

We employed the platform from \cite{edBB2020_AI4EDU}, called edBBplat. It has been designed for capturing data for automatic detection of anomalous behaviors in virtual evaluation environments.

    



\subsection{Sensors}
\label{sensors}
Table \ref{tab:sensors} shows the sensors and the types of data captured by the platform. The data is acquired through a set of activities for the students to complete.


\begin{figure*}[t!]
\centering
\includegraphics[width=\textwidth]{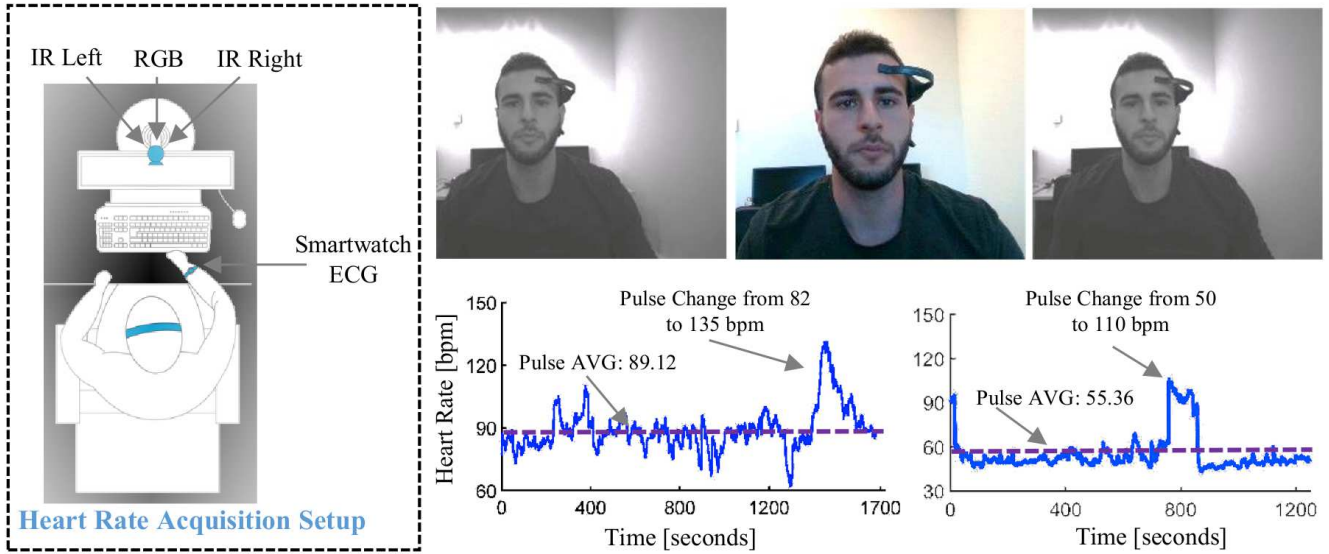} 
\caption{\textbf{Example of the information acquired for heart rate estimation using remote photoplethysmography.} The acquisition setup can be seen in the left diagram. The sensors of the RealSense camera used in this case are the RGB and the left and right near infrared channels (top-right images). We show two different groundtruth heart rates captured with the Huawei Watch 2 smartwatch (bottom-right plots). In these plots, the points in which the users were asked to perform physical activity are highlighted.}
\label{acquisition}
\end{figure*}

The acquisition setup consists of (see Fig.~\ref{acquisition} left):
\begin{itemize}

    \item \textbf{Video}: $4$ RGB cameras ($2$ frontal, $1$ side, and $1$ zenital), $2$ Near Infrared cameras (Intel Real-Sense model D435i), and depth images.

    
    \item \textbf{Pulse and Motion Sensors}: we employed a Huawei Watch 2 smartwatch that captures pulse and motion signals including accelerometer, magnetometer, and gyroscope.



    \item A \textbf{Personal Computer} with Microsoft Windows 10, a mouse, a keyboard, a microphone, and a screen. The computer is used by the students to complete the tasks, while the screen data, the mouse and keyboard dynamics, the audio, and other PC metadata are being acquired in the background. 

\end{itemize}


    


\subsection{Tasks}
\label{tasks}

The activities that conform the platform consist of $8$ different tasks categorized in $3$ main groups:

\begin{itemize}
    \item \textbf{Enrollment form}: meant for obtaining personal data of the students, e.g. name and surname, e-mail address, ID number, and nationality. 
    
    
    \item \textbf{Writing questions}: Since this type of questions are more complex, they can be used to measure the students' cognitive abilities under different situations such as: solving logical problems, describing images, crosswords, finding differences, etc. Additionally, some activities have been designed to induce different emotional states to the participants, e.g. stress or nervousness. 
    
    
    \item \textbf{Multiple choice questions}: These are questions largely used in online assessment platforms and are included to detect the students' attention and focus levels.
    
\end{itemize}



\section{Methods and Challenges}
\label{challenges}


    
    
    
    
    

An example of the employed sensors and of the information that is acquired while a student is completing a task can be seen in Fig.~\ref{acquisition}. The work in \cite{edBB2020_AI4EDU} proposed $5$ different challenges that are relevant to student monitoring:  



 \begin{itemize}
    
    \item \textbf{Challenge $\textbf{1}$ - Attention Estimation:} the estimation of the intensity of mental focus or attention of the students.

    \item \textbf{Challenge $\textbf{2}$ - Anomalous Behavior Detection:} detection of non-allowed activities performed by the students.

    \item \textbf{Challenge $\textbf{3}$ - Performance Prediction:} prediction of accuracy and time necessary for the completion of the tasks.

    \item \textbf{Challenge $\textbf{4}$ - User Authentication:} recognition of the student identity by means of biometric characteristics.
    
\end{itemize}
    
    And, finally, the fifth challenge, which is the main focus of the present paper:
    
 \begin{itemize}
    
    \item \textbf{Challenge $\textbf{5}$ - Pulse Estimation:} changes in the human pulse have showed to be related to altered emotional states and the presence of stress. Emotional states can affect perception and performance. Understanding the emotional state of the student may help in different ways: 1) online adaptation of the session according to the emotional state (e.g. reducing working load and the difficulty or type of the contents); 2) improved performance analysis including emotional features. The objective of this challenge consists in estimating the groundtruth human pulse (obtained from the smartwatch) by using the front camera. Alternatively, the NIR cameras present in the acquisition setup can be used to analyse the potential of this type of sensors.

\end{itemize}

\section{Assessing the Student Heart Rate by rPPG}
\label{estimation}

\begin{figure*}[t]
\centering
\includegraphics[width=\linewidth]{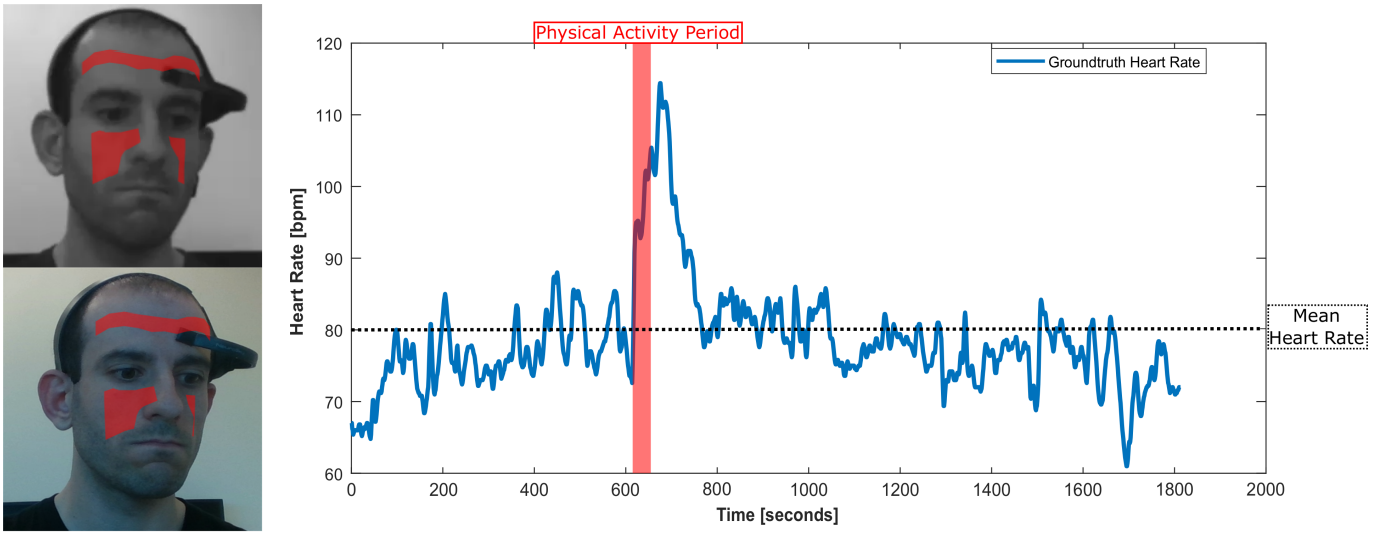} 
\caption{\textbf{Heart Rate estimation using rPPG.} The Regions of Interest for extracting the raw rPPG signals from a NIR image (Left-Up) and the corresponding RGB image (Left-Down) are highlighted in red. Result of the
heart rate estimation (Right). The highest peak in the acquired heart rate corresponds with a moment in which the student was requested to perform a $20$ seconds period of physical activity to get him into an altered state. The mean heart rate for the whole session (sub-challenge 5.1) and the values of the heart rate for $10$ second windows (subchallenge 5.2) are also shown.}
\label{groundtruth}
\end{figure*}

In this study we propose an accurate estimation of the heart rate through Remote Photoplethysmography (rPPG) techniques applied to face biometrics~\cite{REF_HRcomp}. The proposed benchmark is divided into the following two different sub-challenges related to the student activity monitoring:

\begin{itemize}

    \item \textbf{Sub-Challenge 5.1 - Heart Rate averaged by session:} knowing the mean heart rate of a student for a whole session can be useful for comparing these values across different sessions. This way we can track the student's activity along time for detecting unusual events. The average heart rate during the task, the grade obtained, and the student historic data (previous average heart rate and grades) can serve to obtain a detailed picture of the student's performance. 
    
    \item \textbf{Sub-Challenge 5.2 - Heart Rate continuous monitoring:} this challenge consists in dividing each session in shorter temporal windows and estimating the heart rate for each one of them individually. Unlike the first sub-challenge, this approach can be useful for analyzing the state of the student throughout a single session and detecting anomalous behaviors within the session. Additionally, this information is useful to better understand the potential difficulties faced during the tasks.
    
    
\end{itemize}

Plethysmography refers to techniques for measuring the changes in the volume of blood through human vessels. This information can be used to estimate parameters such as heart rate, arterial pressure, blood glucose level, or oxygen saturation levels. The variant called Photoplethysmography (PPG) includes low-cost and noninvasive techniques associated with imagery and the optical properties of the human body \cite{allen2007photoplethysmography}. Oxygenated blood absorbs more light at specific wavelengths than the blood with less oxygen, so measuring over time the amount of light reflected by the tissues of a person, we can estimate his pulse signal and other parameters like respiration variability \cite{shelley2007photoplethysmography}. 

Studies have proven that it is possible to measure the changes in the amount of oxygenated blood through facial video sequences \cite{poh2011advancements}. These techniques are called remote photoplethysmography and their operating principle consists in looking for slight changes in the skin color at video recordings using signal processing methods~\cite{REF_HRcomp}. Remote PPG methods can take advantage of cameras that contain both RGB and near infrared sensors. The NIR spectrum band information is highly invariant to light conditions, providing robustness against this external source of variability at a low cost. The NIR band can also help to derive depth information that could improve the location accuracy of the Regions of Interest (ROI) at face tracking. 

Our approach is based in the one presented in \cite{hernandez2018time} and consists in four main stages: i) we first locate and track $3$ different regions of interest in the student's faces, i.e., the forehead and the right and left cheeks (see Figure~\ref{groundtruth} left); ii) we track the regions during the video and we extract their raw rPPG signals; iii) we postprocess the raw rPPG signals from the $3$ regions using a moving window to isolate the component associated to the pulse by minimizing the other components in the video sequences; and iv) we estimate the value of the heart rate for each temporal window by analyzing the frequency components of the postprocessed rPPG signal and we concatenate all these values for obtaining the heart rate estimation for all the video sequence (see Fig.~\ref{groundtruth} right).


\subsection{Experimental Protocol}



We have acquired $25$ different students while completing the tasks described in Section \ref{tasks}. The duration of each video recording is variable, going from $15$ to $30$ minutes. One session has been recorded for each student. The video sequences have been captured at $30$ frames per second with the Intel RealSense camera (we have used both the RGB and the NIR channels), with a resolution of $1280 \times 720$ pixels. The groundtruth for the heart rate has been acquired with the Huawei Watch 2 smartwatch at a sampling frequency of $1$ Hz. An example of the images captured with the RealSense camera and the smartwatch can be seen in Fig.~\ref{acquisition} right. During the acquisition, each student had to perform physical activity in a different moment of the evaluation in order to put him into an altered state with a higher heart rate. With the physical activity we intended to simulate possible situations in which the pulse of the student may vary due to events such as high stress or cheating attempts. We are aware that physiological changes are highly related with the nature of the stimulus. Changes in the pulse due to physical activity may show different physiological responses that those caused by stress level for example. However, the resulting changes in the heart rate should be similar.   

We decided to use the RGB and the NIR channels in order to compare the results obtained with each type of images. However, in most acquisition setups, the only available sensor will probably be a RGB camera, so we have centered our analysis in the results obtained with that frequency band.

The metric used to report the accuracy in the heart rate estimation challenge is the Mean Average Error (MAE) expressed in beats per minute (bpm). MAE refers to the mean difference in absolute value between the estimated heart rate and the groundtruth. This metric can give us an idea of the average accuracy we can expect of our heart rate estimation method, thus giving us orientation of its possible applications. 

There are slight differences in the protocol we followed for each one of the two sub-challenges. The first step is common to both challenges: we divided the video sequences in temporal windows of a fixed length and we computed a value of the estimated heart rate for each one of these windows. Regarding the groundtruth heart rate, we computed the mean value of the samples acquired with the smartwatch from each temporal window.

\subsubsection{Sub-Challenge 5.1 - Heart Rate averaged by session}
 For computing the mean heart rate of a whole session we calculated the average of the heart rate estimations of all its temporal windows. Then we used the absolute difference between the estimated mean heart rate and the groundtruth as our error metric in beats per minute (bpm). We have selected values for the window length going from $5$ to $20$ seconds with an increment of $5$ seconds.

\subsubsection{Sub-Challenge 5.2 - Heart Rate continuous monitoring}
In this case we took the estimated heart rate and the groundtruth heart rate for each single window and we calculated the absolute difference between them. After that we averaged the error of all the windows inside each video sequence. The results of each session were then combined to produce a single performance measure for the whole dataset, i.e. the Mean Average Error (MAE) expressed in bpm. In this case we explored values for the window length going from $5$ seconds to $20$ seconds with a step of $2$ seconds.



\subsection{Results}
\label{results}


\begin{table}[t]


\caption{Sub-Challenge 5.1: Mean Average Error (in bpm) for heart rate estimation of complete videos}
\label{mean_accuracy}
\begin{center}

\begin{tabular}{ccccc}

\multicolumn{1}{c}{}                                       & \multicolumn{4}{c}{\cellcolor[HTML]{D0D0D0}Temporal Window {(}seconds{)}}  \\ \cline{2-5}

  & \multicolumn{1}{c}{\cellcolor[HTML]{C0C0C0}5} & \multicolumn{1}{c}{\cellcolor[HTML]{C0C0C0}10} & \multicolumn{1}{c}{\cellcolor[HTML]{C0C0C0}15} & \multicolumn{1}{c}{\cellcolor[HTML]{C0C0C0}20}  \\ 
\cline{2-5}

\multicolumn{1}{c|}{\cellcolor[HTML]{D0D0D0}RGB}           & \multicolumn{1}{c|}{10.15}                     & \multicolumn{1}{c|}{5.99}                     & \multicolumn{1}{c|}{6.26}                     & \multicolumn{1}{c|}{6.41}                       \\ 
\cline{2-5}

\multicolumn{1}{c|}{\cellcolor[HTML]{D0D0D0}NIR}           & \multicolumn{1}{c|}{7.62}                          & \multicolumn{1}{c|}{7.13}                          & \multicolumn{1}{c|}{7.08}                          & \multicolumn{1}{c|}{7.40}                              \\ 
\cline{2-5}

\end{tabular}
\end{center}
\end{table}

\begin{table}[t]


\caption{Sub-Challenge 5.2: Mean Average Error (in bpm) for heart rate continuous monitoring}
\label{results_mae}
\begin{center}

\resizebox{\columnwidth}{!}{
\begin{tabular}{cccccccccc}

\multicolumn{1}{c}{}                                       & \multicolumn{9}{c}{\cellcolor[HTML]{D0D0D0}Temporal Window {(}seconds{)}}  \\ 


\multicolumn{1}{c}{}  & \multicolumn{1}{c}{\cellcolor[HTML]{C0C0C0}5} & \multicolumn{1}{c}{\cellcolor[HTML]{C0C0C0}7} & \multicolumn{1}{c}{\cellcolor[HTML]{C0C0C0}9} & \multicolumn{1}{c}{\cellcolor[HTML]{C0C0C0}11} & \multicolumn{1}{c}{\cellcolor[HTML]{C0C0C0}13} & \multicolumn{1}{c}{\cellcolor[HTML]{C0C0C0}15} & \multicolumn{1}{c}{\cellcolor[HTML]{C0C0C0}17}  & \multicolumn{1}{c}{\cellcolor[HTML]{C0C0C0}19}  & \multicolumn{1}{c}{\cellcolor[HTML]{C0C0C0}20} \\ 
\cline{2-10}

\multicolumn{1}{c|}{\cellcolor[HTML]{D0D0D0}RGB}           & \multicolumn{1}{c|}{13.45}                     & \multicolumn{1}{c|}{9.07}                     & \multicolumn{1}{c|}{8.16}                     & \multicolumn{1}{c|}{8.08}                      & \multicolumn{1}{c|}{8.09}                       & \multicolumn{1}{c|}{8.15}                       & \multicolumn{1}{c|}{8.10}  & \multicolumn{1}{c|}{8.19}  & \multicolumn{1}{c|}{8.15} \\ 
\cline{2-10}

\multicolumn{1}{c|}{\cellcolor[HTML]{D0D0D0}NIR}           & \multicolumn{1}{c|}{10.90}                          & \multicolumn{1}{c|}{10.66}                          & \multicolumn{1}{c|}{10.15}                          & \multicolumn{1}{c|}{10.05}                           & \multicolumn{1}{c|}{9.70}                           & \multicolumn{1}{c|}{9.67}                           & \multicolumn{1}{c|}{9.53}  & \multicolumn{1}{c|}{9.63}  & \multicolumn{1}{c|}{9.55}   \\ 
\cline{2-10}

\end{tabular}
}
\end{center}
\end{table}

\subsubsection{Sub-Challenge 5.1 - Heart Rate averaged by session}

In this sub-challenge we have calculated the MAE values for the estimation of the heart rate for complete sessions. The rPPG algorithm used in this work employs information from the three color channels available in RGB videos. However, in NIR videos only one channel is available, so we replicated its information into three different channels to imitate a RGB video. 

In Table \ref{mean_accuracy} we can observe a clear trend of the heart rate estimations, where the NIR videos obtain a higher accuracy when using short video windows, while the RGB-based estimation is the most accurate when using a longer window duration. The accuracy obtained is high for both types of videos, being slightly higher for the NIR band when using short windows, and better for the RGB color channel when using a wider temporal window. We think that this may be caused by the fact that the NIR band is more robust to external illumination changes that affect severely to the rPPG heart rate estimation. However, for longer window sequences, having more information available (three channels instead of one) makes possible to obtain better rPPG signals. 

This sub-challenge may be applicable for monitoring the state of the students between sessions, i.e. knowing in which classes or evaluations the mean heart rate is higher or lower. These alterations may be caused by user impersonation, lack of interest, or a high level of stress.

%


\subsubsection{Sub-Challenge 5.2 - Heart Rate continuous monitoring}

Table \ref{results_mae} shows the performance results for heart rate estimation obtained for different values of the temporal window, going from $5$ seconds to $20$ seconds, and also for both the RGB and the NIR bands. It can be seen that the MAE decreases when increasing the temporal window length because the algorithm has more information for extracting the frequency components correspondent to the heart rate. However, when the window duration reaches a limit (close to $20$ seconds in both cases) the MAE does not further improve due to the variations of the heart rate inside a too long window. Other drawback related to the use of a longer temporal window is the lower temporal resolution of the predictions. If the heart rate changes quickly, a long temporal window will not be able of capturing that behavior. Similarly to the case of the Sub-challenge 5.1, in this case the accuracy is slightly higher for the NIR band when using short windows and better for the RGB color channel when using a wider temporal window.

\begin{figure}[t!]
\centering
\includegraphics[width=1\columnwidth]{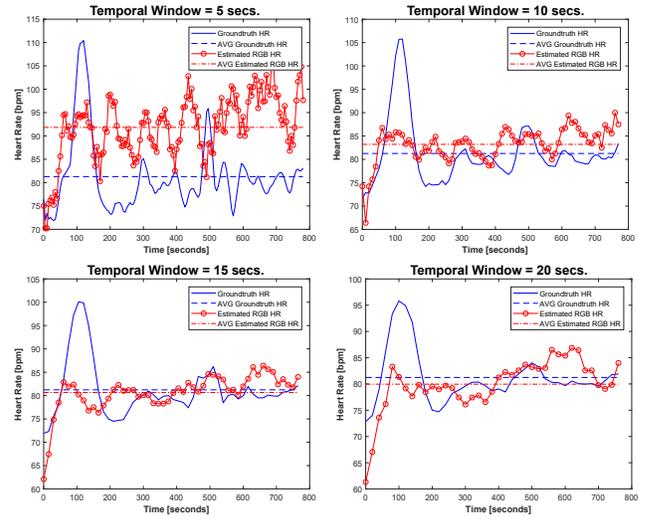} 
\caption{\textbf{Temporal evolution of the heart rate} in a scenario in which the student has been induced to an altered state by means of physical activity at the beginning of the session. The four plots correspond to the same video sequence but with different temporal window lengths. The figure shows the changes in the accuracy when changing the length of the temporal window.}
\label{curves_window}
\end{figure}


Fig.~\ref{curves_window} shows the temporal evolution of the heart rate estimation in a scenario in which the students performed physical activity at some points of the evaluation in order to get their heart rate artificially high. The target is checking if the heart rate estimation algorithm is capable of detecting these changes in the heart rate. 

By inducing alterations we want to simulate a situation in which a student performs any forbidden or inappropriate action, e.g. cheating, that may lead to an altered heart rate. The four plots in the figure correspond to the same video sequence but with a different temporal window length. The figure shows how the estimation algorithm manages to capture the main behavior of the heart rate during the induced alterations. It also reflects the change in the accuracy for the heart rate estimation for the same video sequence when changing the value of the temporal window. 

As has been said previously when commenting the results of Table \ref{results_mae}, a higher value for the temporal window makes the MAE to decrease. This is shown in Fig.~\ref{curves_window} with the plot of the averaged groundtruth and estimated heart rates, that become closer when increasing the temporal window length. However, it can also be seen that even though using smaller windows decreases the general accuracy of the heart rate estimation, it also allows to reflect better the quick changes in the heart rate due to the altered states induced in these experiments. These quick changes in the heart rate can only be captured when using lower values for the temporal window. This way, the decision of what window length must be used depends of the desired application.



\section{Conclusion and Future Work}
\label{conclusion}

In this paper, we have: i) discussed the application of behavioral biometrics for remote education, ii) employed edBBplat \cite{edBB2020_AI4EDU}, a platform of biometrics and behavior for student assessment during virtual education, iii) captured data from sensors that are usually present in remote education (RGB cameras), and also from more advanced sensors like NIR cameras and a smartwatch, and iv) used the acquired NIR and RGB video recordings for estimating the heart rate of the students using rPPG while they are completing a series of virtual evaluation tasks.


The type of information acquired in this work can be used for detecting unusual events during an evaluation task in remote education. Some examples of events that can be detected are: cheating attempts, a stress level out of the ordinary values, drops in the level of attention of the students, or changes in their heart rate.



For future work, we expect to add different types of stimuli that lead to altered states. Correlating those altered states with the information from the other basic and advanced sensors of the platform (EEG band, other cameras, test results, etc.) may be helpful for detecting inappropriate behaviors and other factors such as the stress level, the focus level, or even for trying to predict some variables like the student's performance.


\section{ACKNOWLEDGMENTS}
\label{sec:ack}

This work has been supported by projects: IDEA-FAST (IMI2-2018-15-two-stage-853981), PRIMA (ITN-2019-860315), TRESPASS-ETN (ITN-2019-860813), BIBECA (RTI2018-101248-B-I00 MINECO/FEDER), and edBB (Universidad Autonoma de Madrid, UAM). J. H.-O. is supported by a PhD fellowship from UAM.

\bibliographystyle{IEEEbib}
\bibliography{egbib}

\end{document}